\documentstyle[sprocl,epsfig]{article}

\def\Journal#1#2#3#4{{#1} {\bf #2}, #3 (#4)}

\def\NPB{{\em Nucl. Phys.}\bf B}\def\NPA {{\em Nucl. Phys.}\bf A}
\def\PLB{{\em Phys. Lett.}\bf  B}
\def\PRL{\em Phys. Rev. Lett.}
\def\PRD{{
\em Phys. Rev.}\bf D}\def\PRC{{\em Phys. Rev.}\bf C}

\def\del{\partial}

\def\bfq{{\bf q}}

\def\bfr{{\bf r}}

\def\calO{{\cal O}}\def\calM{{\cal M}}
\def\MeV{\rm MeV}
\def\la{\langle}
\def\be{\begin{equation}}
\def\ee{\end{equation}}
\def\bea{\begin{eqnarray}}
\def\eea{\end{eqnarray}}
\def\bq{\begin{eqnarray}}\def\eq{\end{eqnarray}}
\def\roughly#1{\mathrel{\raise.3ex\hbox{$#1$\kern-.75em%
\lower1ex\hbox{$\sim$}}}}
\def\lsim{\roughly<}
\def\gsim{\roughly>}
\def\Tr{\rm Tr}
\begin{document}
\hfill {KIAS-P98043}
\vskip 1cm

\title{CHIRAL SYMMETRY IN NUCLEAR PHYSICS \footnote{Lectures given at
``The APCTP-RCNP Joint International School on Physics of Hadrons and QCD,"
Osaka, Japan, 12-13 October 1998 and at ``The 1998 YITP Workshop on QCD and
Hadron Physics," Kyoto, Japan, 14-16 October 1998}}

\author{Mannque RHO}

\address{Service de Physique Th\'eorique, CE Saclay\\ 91191
Gif-sur-Yvette, France \footnote{Permanent address}\\ and \\ School
of Physics, Korea Institute for Advanced Study\\ Seoul 130-012,
Korea\\ E-mail: rho@spht.saclay.cea.fr}

\maketitle\abstracts{ How chiral symmetry -- which is a basic
ingredient of quantum chromodynamics (QCD) for light-quark hadrons
-- enters and plays an eminent role in nuclear physics is
discussed. This is done in two steps.  In the first step, I
introduce the notion of the Cheshire Cat Principle which describes
how a strong-interaction phenomenon can be described in various
different languages. In the second step, I treat two cases of
widely different density regimes. The first case deals with
formulating nuclear physics of {\it dilute} nuclear systems with a
focus on both bound and scattering processes in two-nucleon systems
in terms of effective field theories (EFT), e.g., in the
framework of chiral perturbation theory. The second case requires
formulating effective field theories for {\it dense} hadronic
matter such as heavy nuclei and nuclear matter under normal as well
as extreme conditions expected to be encountered in compact stars
and in heavy-ion collisions at relativistic energy. The first case
is rigorously formulated but the second relies on several
assumptions that appear to be reasonable but need ultimately to be
justified. The Cheshire Cat Principle is proposed to play  an essential role in
lending support to the latter development.}

\renewcommand{\thefootnote}{\#\arabic{footnote}}
\setcounter{footnote}{0}

\section{Introduction}\label{introduction}
\indent\indent
Chiral symmetry, now known to be associated with the fact that the
up and down quarks are very light, has been playing an important
role in nuclear physics since a long time, since even before the
advent of QCD~\cite{CS,CND}. In this lecture, I will describe some
of the more recent  developments in this area in which I have
participated.

One of the most notable results obtained in nuclear physics that
are consequences of chiral symmetry is the exchange current
contribution to nuclear responses to electroweak currents. Although
exchange currents are clearly established by theory and experiment,
the role of chiral symmetry -- specifically in terms of the pion
-- is not well appreciated in nuclear physics community. I believe that
this issue marks the turning point from the ``classical'' nuclear
physics to the next generation of hadronic physics probing hadronic
matter under extreme density and/or temperature conditions.

It was pointed out in 1971 by Chemtob and Rho \cite{chemrho} that
the meson-exchange current, a long-standing problem in nuclear
physics, could be efficiently organized in such a way that the main
contribution be calculated by low-energy theorems based on chiral
symmetry. Riska and Brown \cite{riskabrown}, by exploiting the
Chemtob-Rho organization, succeeded to explain quantitatively the
radiative $np$ capture process $n+p\rightarrow d+\gamma$. In 1978,
Kubodera, Delorme and Rho \cite{KDR} proposed what is now called
the ``chiral filter hypothesis" which states that whenever
soft-pion exchanges are {\it un-suppressed} by kinematics or
selection rules, they should provide the dominant model-independent
contribution with the remaining corrections markedly suppressed
while whenever they are {\it suppressed} for the same reason, then
short-distance effects can and do enter at the same level as higher
order corrections, rendering systematic calculations difficult if
not unfeasible. The predictions that followed from this hypothesis
were that the soft-pion effects should be prominent in (1) nuclear
electromagnetic M1 transitions and (2) weak axial-charge
transitions. Both these predictions were unambiguously confirmed by
experiments. The former was confirmed in the large momentum
transfer electrodisintegration of the deuteron $e+d\rightarrow
n+p+e$ performed in 1980's~\cite{frois} and the latter in a series
of experiments and analyses of Warburton in early 1990's
\cite{warburton}. These predictions which were made without direct
recourse to low-energy effective field theories of QCD
 were more recently given a
justification in terms of chiral perturbation theory in 1991
\cite{MR91} \footnote{In the context of chiral perturbation theory,
it is natural that the pion-exchange currents dominate whenever
they are unsuppressed over short-ranged terms although in the
potential both the pion-exchange and contact interactions without
derivatives figure on the same footing. The chiral filter phenomena
in the {\it currents} can be understood by a simple chiral counting
as shown in this paper.} and more rigorously treated in the context
of effective field theories using a cut-off regularization as
described below \cite{PKMR}.

What I believe to be the most crucial aspect in understanding
nuclear phenomena in the context of QCD is that a low-energy
process can be described very accurately in terms of variables that
represent physical (color singlet) objects, that is, the hadron
degrees of freedom that are measured in the laboratory although the
appropriate QCD variables are quarks and gluons. The way this can
be understood from the point of view of QCD is in terms of what is
now called ``Cheshire Cat Principle" (CCP)~\cite{CND} which in a
nutshell can be summarized as an (albeit approximate) equivalence
between the description in terms of hadronic variables and that in
terms of QCD variables for processes that take place in
long-wavelength kinematic regimes. In certain cases,  the CCP can
be exact. For instance, in two space-time dimensions, the
bosonization technique allows one to precisely describe fermionic
excitations in terms of bosonic and vice versa and in the presence
of supersymmetry, there can be exact ``duality,"  a concept which
is becoming very popular in particle physics community
\footnote{In modern string theory \cite{sen}, this is summarized by the
two-fold duality, ``elementality $\leftrightarrow$ composite" and
``classical $\leftrightarrow$ quantum."}. In nonsupersymmetric and
dimension-four world, there is no known transformation that can
give an exact CCP but there is no proof that such a transformation
does not exist. I will show a recent case where in QCD at four
dimensions a fairly good CCP can be seen in nucleon structure.
Another intriguing case, though somewhat different in nature,
 is discussed in \cite{SW}.

At very low energy and low density, nuclear physics can be
accurately described in terms of low-mass hadronic variables with a
cutoff set at the chiral scale given by the vector meson mass,
$\sim 1$ GeV. I will illustrate this in terms of effective field
theories that involve only the nucleon fields and pionic fields. In
fact we will see that even the pion can be integrated out by
setting an effective cutoff at the pion mass and yet a quantitative
agreement can be obtained in the nucleon-only theory. I will focus
on two-nucleon systems  which can be described in terms of a
Lagrangian whose parameters are defined in matter-free space.

I will next jump over the intermediate density regime to  one for
which no systematic chiral perturbation theory approach is
available. This is the case with nuclear matter where new scales
enter due to the presence of a Fermi sea. For this case, two
effective field theories will figure; one, chiral Lagrangian field
theory in a dense background and the other, Fermi-liquid
fixed-point theory. These two by themselves are superb effective
field theories. I will argue that in going to dense hadronic matter
beyond nuclear matter density, the two effective field theories
could be married once Brown-Rho scaling\cite{BR} is introduced. The
arguments used here are more conjectural than rigorous but I will
show a number of evidence that the marriage is a successful one.
\section{Cheshire Cat Principle (CCP)}\label{cheshire}
\indent\indent
When the large $N_C$ property of QCD which implies the skyrmion
picture for baryons and the bag structure of QCD with asymptotic
freedom are properly combined, we have the chiral bag \cite{CND}.
When the topological structure of the skyrmion is incorporated into
the chiral bag, we obtain the (approximate) Cheshire Cat structure.
It has been shown \cite{CND,Hosaka} that at least at low energies,
practically all physical observables, with few exceptions, obey the
Cheshire Cat Principle, in some cases exactly (such as the baryon
charge thanks to topological invariance) and some approximately
(those properties not connected to topology). The reason why the
Cheshire Cat Principle is operative is probably connected to the
existence of something deep in the form of a web of dualities in
gauge theories.

Among the few cases that appeared to deviate from the Cheshire Cat
Principle, a most recent and intriguing one is the flavor singlet
axial charge (FSAC) of the proton, sometimes mistakenly identified
as ``proton spin." A straightforward calculation of the FSAC in the
chiral bag would indicate that the FSAC is small at small bag
radius and grows to order unity at large bag radius which would
mean that the Cheshire Cat is grossly violated. The purpose of this
part of the lecture is to show that if certain vacuum properties of
the bag related to anomalies are properly taken into account, the
CC can be recovered \cite{LPMRV}. Here breaking of chiral symmetry
(i.e., chiral $U_A (1)$) due to quantum effects will be involved.
\subsection{Chiral Bag}
\indent\indent
The most convenient formalism to discuss the CCP is the chiral bag
model (CBM) which is a model that interpolates between the
``bagged'' quark description known as MIT bag and the topological
model, skyrmion, obtainable in the large $N_C$ limit of QCD. This
model consists of the bag of radius $R$ within which the QCD
degrees of freedom, quarks and gluons, live surrounded by the cloud
of pseudo-Goldstone bosons (pions, kaons etc.) and, if needed for
short-distance physics, massive mesons with communication between
the inside and the outside mediated by suitable boundary
conditions. The power of this model is that it can combine both
long-distance (outside) and short-distance (inside) physics
controlled by the boundary terms. Thus it can describe
long-wavelength processes as well as short-wavelength processes
within a single model which is difficult for the skyrmion or the
MIT bag separately. For convenience, one usually uses -- as we
shall do for this lecture -- a spherical geometry for the bag with
a sharp boundary but in principle one could use an arbitrarily
shaped bag with a fuzzy surface. Because of the leakage of various
charges across the boundary, it should not really matter what shape
and how sharp the boundary should be in. The CCP in (octet or
triplet) flavor space has been extensively discussed, e.g., in
\cite{CND}, so I shall not go into that. Let me just focus on the
flavor singlet $U_A (1)$ channel which is of particular interest
because of the chiral anomaly associated with the massive
$\eta^\prime$ degree of freedom. The model we consider has the
$\eta^\prime$ field (which we shall simply call $\eta$)
outside\footnote{The pion field outside plays an important role in
inducing the leakage of the baryon charge. This phenomenon is very
well understood and will be taken into account in the result
although we will not discuss it here.}. This $\eta$ field couples
on the surface to the quarks living inside. The coupling condition
follows from the equation of motion for the quark field and takes
the form
\bea
in^\mu \gamma_\mu\psi=e^{i\frac{\eta}{f}\gamma_5}\psi (R)
\label{eoms}
\eea
where $f$ is a mass-dimension-1 constant related to the $\eta$
decay constant and $n^\mu$ is a unit normal to the surface. This
condition assures at the classical level that the baryon current is
conserved \footnote{We know however that the hedgehog pions induce
a violation of the baryon current but the $\eta$ field does not
participate in this violation since there is no topological
component associated with this pseudoscalar meson.}

When the $U_A (1)$ field is present, it can also couple to the
gluons living inside. That gives a boundary condition that involves
the $\eta$ field and the gluon field. To deduce this condition,
consider the normal components of the FSAC on both sides of the
boundary. Coming from outside, one has
\bea
f n^\mu \del_\mu \eta (R).
\eea
This is well-defined at the surface since it is a nice local operator. Now
coming from inside, we have
\bea
\frac 12 n^\mu\bar{\psi}\gamma_\mu\gamma_5\psi (R).\label{naive}
\eea
This however is not well-defined since the bilinear in fermion
fields at the same space-time point is singular. Regularizing it by
point-splitting a la Schwinger reveals that the regularized form of
(\ref{naive}) has an additional term involving the Chern-Simons
current
 $K^\mu=\epsilon^{\mu\nu\alpha\beta} (G_\nu^a
G_{\alpha\beta}^a -\frac 23 f^{abc} g G_\nu^a G_\alpha^b
G_\beta^c)$ given in terms of the color gauge field $G^a_\mu$,
\bea
\frac 12 \hat{n}\cdot({\bar{\psi}}\vec{ \gamma}\gamma_5\psi)_{\tiny NO}
 + \frac{N_F g^2}{16\pi^2 } \hat{n}\cdot K\label{bc}
\eea
where the subscript ``NO'' stands for normal ordering. Now the
Chern-Simons current on the surface is not gauge-invariant, so
(\ref{bc}) is not gauge-invariant. This violation of gauge
invariance, which is not acceptable, can be understood as ``color
anomaly.'' The color anomaly arises as a consequence of the
celebrated chiral anomaly associated with the $U_A (1)$ symmetry
which is broken due to quantum fluctuations in the vacuum. The
flavor singlet axial current inside has the anomaly and hence the
color leaks out as a consequence. To prevent the leakage of the
color, the simplest (though perhaps not the only) way is to put a
boundary condition that ``soaks up'' this leaking color charge.
It has been shown \cite{CND} that the following surface counter
term does the job:
\bea {\cal
L}_{CT}=i\frac{g^2}{32\pi^2}\oint_{\Sigma} d\beta K^\mu n_\mu
({\Tr}{ \ln} U^\dagger -{\Tr} {\ln} U)\label{lct}
\eea
where $N_F$
is the number of flavors (here taken to be =3), $\beta$ is a point
on a surface $\Sigma$, $n^\mu$ is the outward normal to the bag
surface, $U$ is the $U(N_F)$ matrix-valued field written as
$U=e^{i\pi/f} e^{i\eta/f}$.
Note that (\ref{lct}) manifestly breaks color gauge invariance
(both large and small, the latter due to the bag), so the action
of the chiral bag model with this term is not gauge invariant at
the classical level but as shown in \cite{nrwz}, when quantum
fluctuations are calculated, there appears an induced anomaly term
on the surface which exactly cancels this term. Thus gauge
invariance is restored at the quantum level. Presence of the boundary
term (\ref{lct}) implies that the color electric and magnetic fields must
satisfy boundary conditions that have coupling to the $\eta$ field,
\bea
\hat{n}\cdot \vec{E}^a=-\frac{N_F g^2}{8\pi^2 f} \hat{n}\cdot \vec{B}^a
\eta,\label{E}
\eea
\bea
\hat{n}\times \vec{B}^a=\frac{N_F g^2}{8\pi^2 f} \hat{n}\times \vec{E}^a
\eta. \label{B}
\eea

These together with the ``matter'' boundary conditions (\ref{eoms})
and the axial current continuity
\bea
\frac 12 \hat{n}\cdot({\bar{\psi}}\vec{ \gamma}\gamma_5\psi)_{\tiny NO}(R)
 =f n^\mu \del_\mu \eta (R)\label{matterbc}
\eea
define the surface boundary conditions of our CBM to be satisfied
by the usual equations of motion of the quarks and gluons inside
the bag and of the massive $\eta$ field outside of the bag.
\subsection{Flavor-Singlet Axial Charge (FSAC)}
\indent\indent
We now study the flavor singlet axial charge of the proton. We define the
axial current as
\bea
A^\mu =A^\mu_B \Theta_B + A^\mu_M \Theta_M.\label{current}
\eea
Since we will be dealing only with the flavor-singlet axial
current, we will omit the flavor index in the current. We shall use
the short-hand notations $\Theta_B=\theta (R-r)$ and
$\Theta_M=\theta (r-R)$ with $R$ being the radius of the bag. We
demand that the $U_A (1)$ anomaly be given in this model by
\bq
\partial_\mu A^\mu =
\frac{\alpha_s N_F}{2\pi}\sum_a \vec{E}^a \cdot \vec{B}^a \Theta_{B}+
f m_\eta^2 \eta \Theta_{M}.\label{abj}
\eq
Our task is to construct the FSAC in the chiral bag model that is
gauge-invariant and consistent with this anomaly equation. Our
basic assumption is that in the nonperturbative sector outside of
the bag, the only relevant $U_A (1)$ degree of freedom is the
massive $\eta^\prime$ field. This assumption allows us to write
\bq
A^\mu_M = A^\mu_\eta = f\del^\mu \eta
\eq
with the divergence
\bq
\del_\mu A^\mu_\eta = fm_\eta^2 \eta.
\eq
Now the question is: what is the gauge-invariant and regularized
$A^\mu_B$ such that the anomaly (\ref{abj}) is satisfied? To
address this question, we rewrite the current (\ref{current}) as
\bq
A^\mu=A_{B_{Q}}^\mu + A_{B_{G}}^\mu + A_\eta^\mu\label{sep}
\eq
such that
\bea
\partial_\mu (A_{B_{Q}}^\mu + A_\eta^\mu) &=& f m_\eta^2 \eta
\Theta_{M},\label{Dbag}\\
\partial_\mu A_{B_G}^\mu &=&
\frac{\alpha_s N_F}{2\pi}\sum_a \vec{E}^a \cdot \vec{B}^a
\Theta_{B}. \label{Dmeson} \eea The subindices Q and G imply that
these currents are written in terms of quark and gluon fields
respectively. In writing (\ref{Dbag}), we have ignored the up and
down quark masses.

\begin{figure}
%\vskip 2ex
\centerline{\epsfig{file=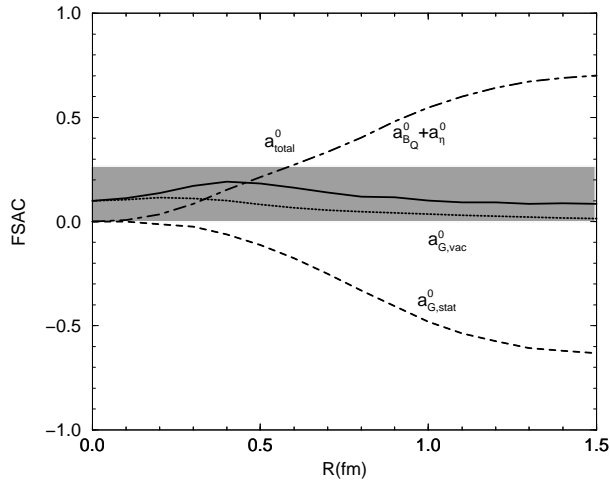, width=9cm}}
\caption{Various contributions to the flavor singlet axial current
of the proton as a function of bag radius and comparison with the
experiment: (a) ``matter'' (quark plus $\eta$) contribution ($a^0_{B_Q} +
a^0_\eta$), (b) the contribution of the static gluons due to quark
source ($a^0_{G,stat}$) , (c) the gluon vacuum contribution
($a^0_{G,vac}$), and (d) their sum ($a^0_{total})$. The shaded area
corresponds to the range implied by experiments.}\label{fig1}
\end{figure}

The matrix element of the ``matter'' currents satisfying (\ref{Dbag}) is
quite simple. The quark current inside the bag is
\bq
 A^\mu_{B_Q} = \bar{\Psi} \gamma^\mu \gamma_5 \Psi
\eq
where $\Psi$ should be understood to be the {\it bagged} quark
field. The proton matrix element
\bq a^0_{B_Q} \equiv \langle p|\int_B d^3r \bar{\Psi} \gamma_3
\gamma_5 \Psi|p\rangle \label{aq}
\eq
can be computed readily taking into account of the leakage of the baryon
charge
due to the hedgehog pion we know how to compute. As for the
$\eta$ contribution, one sees that the boundary condition (\ref{matterbc})
relates it directly to the quark contribution. A simple calculation gives
\bq
 a^0_\eta \equiv \frac{1 +y_\eta}{2(1+y_\eta)
+y_\eta^2} \langle p|\int_B d^3r \bar{\Psi} \gamma_3 \gamma_5
\Psi|p\rangle \label{aeta}
\eq
with $y_\eta=m_\eta R$. The first crucial observation at this point
is that the matter contribution, the sum of (\ref{aq}) and
(\ref{aeta}), will vanish as the radius is shrunk to zero and will
increase monotonically as the bag is increased. This can be seen in
Fig.\ref{fig1}. This feature by itself would show a violent
breakdown of the CCP.

We now turn to the gluon contribution which is a lot more difficult to
calculate accurately. Only an approximate calculation exists at the moment
but I believe it is qualitatively correct.

To see what is involved, we first write down the relevant matrix element
which follows from the relation (\ref{Dmeson}),
 \bq
a^0_{G} = \langle p|-\frac{N_F\alpha_s}{\pi} \int_B d^3r x_3
\vec{E}^a \cdot \vec{B}^a|p\rangle.
\label{ag}\eq
It is not obvious but this matrix element can be shown to be gauge-invariant.
Since we are dealing with a cavity system {\it and} the color anomaly which
requires coupling to the $\eta$ field in the gluon boundary conditions
(\ref{E}) and (\ref{B}), we have a double Casimir problem here. Only when
the Casimir contributions are properly taken into account can we recover
a fully consistent gauge-invariant result.
Doing this calculation in full generality is not going to be an easy
matter~\footnote{Such a calculation is in progress at the time of this
writing.}, so we shall proceed less rigorously.

The gluon contribution can be roughly divided into two parts: One is the
(presumably dominant) effect coming from the quark and $\eta$ sources and
the other is the contribution coming from a vacuum change due to the $\eta$
coupling. The first involves the valence quarks and static gluon fields.
The calculation for this was already made in previous papers (references
are given in \cite{LPMRV}) and only a slight
extension including the $\eta$ field in the boundary condition yields the
desired result. The result shown in Fig.\ref{fig1} as $a^0_{G,stat}$ provides
us with the second important information and that is that the dominant
gluon contribution comes with a negative sign, so
canceling nearly completely the matter contribution.
It seems reasonable as a first approximation
to conclude that $a^0_{B_Q} +
a^0_\eta+a^0_{G,stat}\approx 0$. What is left must therefore be the Casimir
contribution.

The simplest way to compute the Casimir effect is to assume CCP and
check that the assumption is verified a posteriori. In \cite{nrwz},
it is shown that the CCP allows us to write -- via the color
boundary conditions (\ref{E}) and (\ref{B}) -- the following local
operator
\begin{equation}
\vec{ E}^a \cdot \vec{ B}^a (x)\approx -\frac{N_F g^2}{8\pi^2}
\frac{\eta (R)}{f} \frac12 G^2 (x). \label{localform}\end{equation}
Here only
the term up to the first order in $\eta$ is retained in the
right-hand side. Applied to the CBM,
the couplings are to be understood as the average bag
couplings and the gluon fields are to be expressed in the {\it cavity
vacuum} through a mode expansion. In fact, by comparing the
expression for the $\eta^\prime$ mass derived in \cite{nrwz} using
Eq.(\ref{localform}) with that obtained by Novikov et al \cite{NSVZ} in a
QCD sum-rule method, we note that the matrix element of the $G^2$ in
(\ref{localform}) should be evaluated in the {\it absence} of light
quarks. This means, in the bag model, the {\it cavity vacuum}.
The corresponding
Casimir calculation involves a mode sum regularized, say,
by the heat-kernel method.
There is some subtlety involved in this calculation but we will not
go into details here (see \cite{LPMRV} for details).
The results are summarized in
Fig.\ref{fig1} compared with the presently accepted range of the experimental
values obtained in deep inelastic lepton scattering from the nucleon.
We see that the CCP is recovered via the Casimir effect.

To summarize, the CCP holds fairly well in {\it all} low-energy
properties of baryon dynamics. More tantalizingly, there is a hint for a
one-to-one mapping between quark-gluon matter at high density and hadron
matter at low density \cite{SW}. We will later argue that the CCP
holds in normal nuclear matter as well as in high density matter before the
phase transition.
\section{Effective Field Theories (EFT) For Nuclei}
\subsection{The Essence of EFT}
\indent\indent
The idea of effective quantum field theory\cite{effective,pol,lepageTASI}
is extremely simple. At low energy where non-perturbative effects of
QCD dominate, the relevant degrees of freedom are not quarks and
gluons but hadrons. We do not have to know how the quark-gluon variables
get transformed into hadron variables. We may simply assume the CCP
described above as the mechanism for the transformation.
Let the relevant hadronic degrees of freedom be
represented by the generic field $\Phi$. Separate the field into
two parts; the low-energy part $\Phi_L$ in which we are interested
and the high-energy part $\Phi_H$ which does not interest us
directly. We delineate the two parts at the momentum scale characterized
by a cutoff, say, $\Lambda_1$.
In the generating functional $Z$
that we want to compute, integrate out the $\Phi_H$ field and
define an effective action $S^{eff}_{\Lambda_1}[\Phi_L] =\frac{1}{i}\ln
\left(\int [d\Phi_H]e^{iS[\Phi_L,\Phi_H]}\right)$.
Then what we need to compute is
\bea
Z=\int_{\Lambda_1} [d\Phi_L] e^{iS^{eff}_{\Lambda_1} [\Phi_L]}.
\eea
Given the right degrees of freedom effective below the cutoff
$\Lambda_1$, we expand the effective action as
$S^{eff}_{\Lambda_1}[\Phi_L]=\sum_i^\infty C_i Q_i[\Phi_L]$,
where $Q_i[\Phi_L]$ are local field operators allowed by the symmetries
of the problem,
and $C_i$'s are constants,
which are required to satisfy  the ``naturalness condition".
The $Q_i[\Phi_L]$'s are ordered in such a manner that,
for low-energy processes,
the importance of the $Q_i[\Phi_L]$ term diminishes as $i$ increases.
The contribution of the $\Phi_H$ degrees of freedom
integrated out from the action is not simply discarded but instead gets
lodged in the coefficients $C_i$  as well as in higher-dimensional operators.
The strategy of effective field theories is to truncate the series
at a manageable finite order, and obtain desired accuracy to capture the
essence of the physics involved.
The separation of $\Phi$ into $\Phi_L$ and $\Phi_H$
involves two steps: The first is to ``decimate'' the degrees of freedom,
that is, to choose the fields $\Phi_L$
relevant for the physics to be incorporated explicitly into the Lagrangian,
and the second is a regularization procedure which
regulates the high-momentum contribution.
Let $m$ be the lightest mass of the degrees of freedom
that are integrated out.
Clearly the most physically transparent regularization
is to introduce a momentum cutoff with $\Lambda_1 \sim m$.
A different cutoff introduces different coefficients $C_i$ for the
given set of fields.
The idea is then to pick a value of the cutoff
that is low enough to avoid unnecessary complications
but high enough to avoid throwing away relevant physics.

This defines an effective field theory. The rest is just
prescription and work. Technical details may differ, some more
convenient and more elegant than others, but when done well, they
should all give equally good results.

In the full (untruncated) theory, the location of the cutoff is
entirely arbitrary, so we could have chosen any values of the
cutoff other than $\Lambda_1$ without changing physics. The cut-off
dependence of $Q_i[\Phi_L]$ is compensated by the corresponding
changes in $C_i$ through the renormalization group equation. In
such a theory, which is rarely available in four dimensions,
physical observables should be strictly independent of where the
cutoff is set. This of course will not be the case when the series
is truncated. However, for a suitably truncated effective theory to
be predictive, the observables should be more or less insensitive
to where the cutoff is put. Should there be strong cutoff
dependence, it would signal that there is something amiss in the
scheme. For instance, it could be a signal for the presence of new
degrees of freedom or of ``new physics" that must be incorporated
and/or for the necessity of including higher-order terms.
%How  ``new physics" enters in our case will be studied in terms of the role
%that the pion -- a ``new degree of freedom" -- plays at very low energy.
%Here we of course know exactly what the ``new physics" is but
%there will be a lesson to learn from this exercise in exploring beyond
%``standard models". The following ``toy model" illustrates our point.
%Let us imagine that we knew that QCD is the correct microscopic theory for
%the strong interactions with chiral symmetry realized in the Nambu-Goldstone
%mode but the pions were not  yet discovered and that the only ``effective"
%degrees of freedom we had at hand were the proton and the neutron.
%Now the question to ask is how to set up a systematic theory that could
%describe very low-energy properties of the deuteron, the NN scattering
%and the electroweak matrix elements and in doing so how to infer that a
%``new degree of freedom" (pion) lurked at $\sim 140$ MeV.
%We will see in Section 6 how this question could be studied
%in the real low-energy world of nuclei.

An important aspect of effective field theory is that the result
should not depend upon what sort of regularization one uses. In
principle, therefore, one could equally well use the dimensional
regularization or the cut-off regularization or something else as
long as one makes sure that all relevant symmetries are properly
implemented. Some of these issues in the context of the dimensional
regularization and its variants were addressed recently in
\cite{gegelia2}. Here we will use the cut-off regularization for
the reason that the procedure is simple and physically
transparent~\cite{lepage} when limited to the order we shall adopt,
namely, to the next-to-leading order (NLO). For some of the
problems treated here such as neutron-proton ($np$) scattering, one
could also use in a predictive manner a modified dimensional
regularization in which certain power divergences are removed
(i.e., PDS scheme)\cite{KSWPDS}.
\subsection{An EFT At Work}
\indent\indent
We shall apply the above strategy to two-nucleon systems at very
low energy. Suppose we look at processes that involve momentum
$p\ll m_\pi\approx 140$ MeV. We can approach the problem in two
ways. Since the momentum involved is much less than the pion mass,
we can simply integrate out the pion and treat the nucleon field
only. This means that the physics of the pion will appear in the
coefficients of the terms we write down in the Lagrangian that we
wish to solve. Next we will restore the pion and put it explicitly
into the calculation. The coefficients of the terms in the
Lagrangian will now be different from the case when the pion was
integrated out. We will compare the two cases and learn how ``new
physics'' in the form of the pion emerges~\cite{pmr,pkmr-prc}. 
As mentioned, if done
correctly, it should not matter which regularization one uses. Here
we shall use the cut-off regularization which is now known to be
the simplest and trouble-free regularization in the market. There
have been lots of papers written in connection with the $np$
scattering at low energy discussing the merits and demerits of
various regularization schemes. Readers interested in what the hot
debates are all about could find them in \cite{debate}. The issue
is quite interesting theoretically but at the end of the day, it is
fair to say that using the cut-off regularization is still the
physically most transparent (although perhaps not the most elegant)
way to do the calculation. I won't go into this matter here which
is in some sense academic. I could also spend all my lecture time
discussing scattering only but I won't do this either. Instead I
will treat {\it all} two-nucleon problems at the same time,
including $np$ and $pp$ channels, bound states, scattering states
as well as electroweak transitions involving the deuteron and
unbound states. Work along this line with a cutoff regularization
was done previously by Ord\'o\~nez et 
al~\cite{ordonezetal}.
\subsubsection{$\bullet$ Counting rule}
\indent\indent
Since we are going to use the cut-off regularization, Weinberg's original
chiral counting rule is perfectly fine. This will deviate from the PDS scheme
in treating the pion when pions figure in the theory. As I shall mention later,
Weinberg's counting scheme has a predictive power in the form of chiral filter
which the PDS scheme lacks. More on this later.

Start with the case where the pion is present.
Let $Q$ be the typical momentum scale of the process or the pion mass,
which is regarded as ``small" compared to the chiral scale
$\Lambda_\chi \sim 1$ GeV.
The counting rule given by Weinberg\cite{weinberg} is that,
a Feynman graph comprised of $A$ nucleons,
$N_E$ external (electroweak) fields,
$L$ {\it irreducible} loops and $C$-separated pieces
is order of $\calO(Q^\nu)$ with
\bea
\mbox{ChPT:}\ \ \
\nu &=& 2 L + 2 (C-1) + 2 - (A+N_E) + \sum_i \bar \nu_i,
\label{nuChPT}\\
\bar \nu_i &\equiv& d_i + \frac{n_i}{2} + e_i -2
\label{barnu}\eea
where we have characterized each vertex $i$
by $d_i$ the number of derivatives and/or the power of the pion
mass, $n_i$ that of nucleon fields and $e_i$ that of external
(electroweak) fields. The quantities $\bar \nu_i$ and $C$ are
defined so that $\bar \nu_i\ge 0$ even in the presence of external
fields \cite{MR91} and $(C-1)=0$ for connected diagrams.

Now if the pion is integrated out, there are no irreducible loops
and the corresponding counting rule can be
obtained simply by putting the number of loops $L$ equal to zero,
\bea
\mbox{EFT:}\ \ \
\nu = 2 (C-1) + 2 - (A+N_E) + \sum_i \bar \nu_i
\eea
with the same definition for $\bar \nu_i$ given in eq.(\ref{barnu}).
But the meaning of $d_i$ in eq.(\ref{barnu}) is changed: Since the
pion field is integrated out, the $d_i$ stands here only for the
number of derivatives, and not the power of the pion mass. Then, up
to the next-to-leading order (NLO),
the resulting irreducible vertex or
potential consists of contact interactions
and  two spatial derivatives thereof.

Electromagnetic interactions have different counting rules
and should be treated separately.
Since the Coulomb interaction (between protons) is given as
$\frac{\alpha}{\bfq^2}$ (where $\alpha\simeq 1/137$ is the fine-structure
constant and $\bfq$ is the momentum transferred),
it is of relevance in an extremely small momentum region
while it becomes irrelevant in other regions due to
the smallness of $\alpha$.
Since we will be  interested also in the proton fusion at
threshold, we will explicitly include the Coulomb potential
in the $pp$ sector.
\subsubsection{$\bullet$ Putting the cutoff}
\indent\indent
To the order we consider here for the irreducible graphs,
that is, to NLO, all we need is to
put the cutoff on the irreducible vertex (call potential from now on).
Given a potential in the momentum space ${\cal V}$ which we are to iterate to
infinite order in the {\it reducible} channel to get the bound  or quasi-bound
state, the regularized
potential in coordinate space is taken as
\bea
V(\bfr) &\equiv& \int\! \frac{d^3\bfq}{(2\pi)^3}
\, \ e^{i\bfq\cdot \bfr}\, S_\Lambda(\bfq^2)\,
 {\cal V}(\bfq),
\label{regV}\eea
where $S_\Lambda(\bfq^2)$ is the regulator with a cutoff $\Lambda$.
For our purpose it is convenient to take the Gaussian
regulator
\be
S_\Lambda(\bfq^2) = \exp\left( - \frac{\bfq^2}{2 \Lambda^2} \right).
\label{Sdef}\ee
Summing the infinite diagrams with the potential is equivalent to solving
Schr\"odinger equation with the potential $V(\bfr)$. Since in our
scheme we should not let $\Lambda$ become too large,
the Schr\"odinger equation is well-defined.
\subsubsection{$\bullet$ Potential without the pion}
\indent\indent
In the absence of the pion, the potential to NLO is simply
\bea
{\cal V}_{no-\pi}(\bfq) &=&
%- \tau_1\cdot\tau_2 \,\frac{g_A^2}{4 f_\pi^2}\,
%\frac{ \sigma_1\cdot \bfq\,\sigma_2\cdot \bfq}{\bfq^2+m_\pi^2}
%+
\frac{4\pi}{M} \left[C_0 + (C_2 \delta ^{ij} + D_2 \sigma^{ij})
 q^i q^j \right]
+ Z_1 Z_2 \frac{\alpha}{\bfq^2}
\label{Vq}\eea
with
\begin{equation}
\sigma^{ij} = \frac{3}{\sqrt{8}} \left(
\frac{\sigma_1^i \sigma_2^j + \sigma_1^j \sigma_2^i}{2}
- \frac{\delta^{ij}}{3} \sigma_1 \cdot \sigma_2 \right),
\end{equation}
where $\bfq$ is the momentum transferred,
% $g_A \simeq 1.26$ the
%axial-vector coupling constant, $f_\pi\simeq 93\ \mbox{MeV}$ the
%pion decay constant,
$M\simeq 940\ \MeV$ the nucleon mass, and $Z_1 Z_2=1$ for the $pp$
channel and zero otherwise. For the proton fusion process, we will
also consider ${\cal O}(\alpha^2)$ corrections, i.e., the vacuum
polarization (VP) potential and the two-photon-exchange (C2)
potential; as in \cite{pkmr-apj}, these will be treated
perturbatively. The parameters $C$'s and $D_2$ are defined for each
channel. The $D_2$ term, however, is effective only for the
spin-triplet channel. Thus, there are two parameters ($C_0$ and
$C_2$) for each of the $pp$ and $np$ ${}^1S_0$ channels, and three
($C_0$, $C_2$ and $D_2$) for the ${}^3S_1$ channel. These
parameters will be determined from scattering and bound-state
experimental data.
\subsubsection{$\bullet$ Potential with the pion}
\indent\indent
To the order considered, the contribution of the pion is simple
and parameter-free. There are no loops in the irreducible channel, so
all we need is the potential
\bea
{\cal V}(\bfq) =
- \tau_1\cdot\tau_2 \,\frac{g_A^2}{4 f_\pi^2}\,
\frac{ \sigma_1\cdot \bfq\,\sigma_2\cdot \bfq}{\bfq^2+m_\pi^2}
+{\cal V}_{no-\pi}(\bfq)\label{Vqp}\eea where  $g_A \simeq 1.26$ is
the axial-vector coupling constant and $f_\pi\simeq 93\ \mbox{MeV}$
is the pion decay constant.
\subsubsection{$\bullet$ Renormalization}
\indent\indent
Since the cutoff is finite, there are no divergences, so for a
given $\Lambda$, the renormalization amounts to determining the
constants that appear in the potential. All the constants are fixed
in our procedure by the scattering lengths and effective ranges for
$np$ and $pp$ in $^1S_0$ channel and the binding energy of the
deuteron ($B_d$), the deuteron $D/S$ ratio ($\eta_d$) and the wave
function normalization factor ($A_s$), all of which are given very
accurately by experiments. For the $pp$ channel, one has to take
into account the Coulomb and other electromagnetic effects. This
can be done in various ways with little ambiguity~\cite{ravndal}.
\subsubsection{$\bullet$ Observables}
\indent\indent
The following observables will be {\it predicted} and compared
with experiments.
\begin{enumerate}
\item Phase shifts for $np$ $^1S_0$ channel.
\item The charge radius $r_d$, quadrupole moment $Q_d$, D-state probability
$p_D$ and magnetic moment $\mu_d$ of the deuteron.
\item The radiative $np$ capture $n+p\rightarrow d+\gamma$.
\item The solar fusion process $p+p\rightarrow d+e^+ +\nu_e$.
\end{enumerate}
\subsection{The Power of EFT: Comparison with Experiments}
\indent\indent
We will now show that the theory works impressively. We will also
see the power of the chiral filter in our scheme. The deuteron
properties and electroweak matrix elements will be treated {\it
with} and {\it without} the pion. The scattering phase shifts will
be shown later.
\subsubsection{$\bullet$ Results without the pion}
\indent\indent
Without the pion, we expect that the cutoff should be around the
pion mass $\sim 140$ MeV. The results are listed in Table
\ref{table1}. They are remarkably stable against the cutoff as long
as it is near or above the pion mass. The agreement with experiment
is also quite impressive.
\begin{table}[tbp]
\caption{\protect \small
The next-to-leading order (NLO) results {\em without} the pion
field. The static properties of the deuteron, the M1 and GT
amplitudes are listed for various choices of the cutoff $\Lambda$.
The low-energy input parameters are the scattering lengths and
effective ranges for the $np$ and $pp$ ${}^1S_0$ channels, and
$B_d$, $A_s$ and $\eta_d$ for the deuteron channel.}
\label{table1}
\begin{center}
\begin{tabular}{|c|cccccc|cc|} \hline
$\Lambda$ (MeV) & $140$ & $150$ & $175$ & $200$ & $225$ & $250$ &
 Exp. & $v_{18}$\cite{v18} \\
\hline \hline
$r_d$ (fm)
   & 1.973 & 1.972 & 1.974 & 1.978 & 1.983 & 1.987 & 1.966(7) & 1.967 \\
$Q_d$ ($\mbox{fm}^2$)
  & 0.259 & 0.268 & 0.287 & 0.302 & 0.312 & 0.319 & 0.286 & 0.270 \\
$P_D$ (\%)
  & 2.32  & 2.83 & 4.34 & 6.14 & 8.09 & 9.90 & $-$ & 5.76 \\
$\mu_d$
  & 0.867 & 0.864 & 0.855  & 0.845 & 0.834 & 0.823  & 0.8574 & 0.847 \\
\hline
${\cal M}_{\rm M1}$ (fm)
   & 3.995 & 3.989 & 3.973 & 3.955 & 3.936 & 3.918 & $-$ & 3.979 \\
\hline
${\cal M}_{\rm GT}$ (fm)
  &4.887 & 4.881 & 4.864 & 4.846 & 4.827 & 4.810 & $-$ & 4.859 \\
\hline
\end{tabular} \end{center} \end{table}
\subsubsection{$\bullet$ Results with the pion}
\indent\indent
With the pion figuring explicitly, we expect the cutoff to be at
the next mass scale $\gsim 2m_\pi\sim 300$ MeV. The results are
given in Table \ref{table2}. The results are even more spectacular.
\begin{table}[tbp]
\caption{\small
The NLO results {\em with} pion field. See the caption of Table
\ref{table1}.}
\label{table2}
\begin{center}
\begin{tabular}{|c|cccccc|cc|} \hline
$\Lambda$ (MeV) & $200$ & 250 & $300$ & 350 & $400$ & $500$ &
 Exp. & $v_{18}$\cite{v18} \\
\hline \hline
$r_d$ (fm)
   & 1.963 & 1.965 & 1.966 & 1.967 & 1.968 & 1.968 & 1.966(7) & 1.967 \\
$Q_d$ ($\mbox{fm}^2$)
   & 0.261 & 0.268 & 0.272 & 0.273 & 0.274 & 0.274 & 0.286 & 0.270 \\
$P_D$ (\%)
   & 3.16 & 4.11 & 4.77 & 5.16 & 5.35 & 5.39 & $-$ & 5.76 \\
$\mu_d$
   & 0.862 & 0.856 & 0.853 & 0.850 & 0.849 & 0.849 & 0.857 & 0.847 \\
\hline
${\cal M}_{\rm M1}$ (fm)
   & 3.987 & 3.976 & 3.968 & 3.963 & 3.958 & 3.952 & $-$ & 3.979 \\
\hline
${\cal M}_{\rm GT}$ (fm)
 & 4.884 & 4.874 & 4.867 & 4.862 & 4.859 & 4.854 & $-$ & 4.859 \\
\hline
\end{tabular} \end{center} \end{table}

\subsubsection{$\bullet$ Meson-exchange current corrections}
\indent\indent
In order to complete the calculations for the $np$ capture and the
proton fusion process, the exchange-current corrections should be
calculated.

For the M1 transition for the $np$ capture, the leading correction
to the dominant single-particle matrix element comes at the next
order from one-soft-pion exchange. This is in accordance with the
chiral filter argument. A detailed bench-mark calculation in ChPT
\cite{pmr} that includes loop corrections (NNLO relative to the
leading M1 matrix element) finds 9.29\% correction to the cross
section for the radiative $np$ capture getting $\sigma^{th}=334\pm
3\ {\rm mb}$ in agreement with the experiment
$\sigma^{exp}=334.2\pm0.5\ {\rm mb}$, the 1\% uncertainty in the 
theoretical result being due
to short-distance physics incompletely accounted for in chiral
perturbation theory.

The meson-exchange correction to the Gamow-Teller matrix is a
different story since it is {\it not} protected by the chiral
filter. In order to assure an accuracy, one would have to go to
${\cal O} (Q^n)$ with $n\geq 3$. So far, calculations have been done
up to ${\cal O} (Q^3)$ but there is no reason to believe that terms
of order with $n\geq 4$ are negligible compared with what has been
calculated. The predicted $S$ factor for the solar burning process
\cite{pkmr-apj} comes out to be
 $S^{ChPT}=4.05 \times 10^{-25}\ {\rm MeV-barn}$ with
a possible uncertainty of $\sim 2\%$ due to higher-order corrections
in the meson-exchange currents that have not been evaluated yet. 
There is no direct check of
this prediction. It is however quite close to what has been
obtained by the usual potential model calculation (with realistic
potentials), $S^{pot}=4.00\times 10^{-25} (1\pm 0.007^{+0.020}_{-0.011})
\ {\rm MeV-barn}$. As is well known now, this leads to the celebrated solar
neutrino problem within the Standard Model. The neutrino mass
resolution to the problem is a hot topic lying in a different
domain of physics.

\subsubsection{$\bullet$ $np$ $^1S_0$ phase shifts}
\indent\indent
Because of the large scattering length involved in this channel,
naive dimensionally regularized EFTs do not work in a natural way.
Much ``sweat and blood' have been shed on the issue of how to
rescue the dimensional regularization scheme from demise. The final
rescue was found in the PDS scheme \cite{KSWPDS} with a slight
modification of the counting rule together with the demotion of the
pion to a sub-dominant role. All this technical problem is avoided
in a straightforward way in the cut-off regularization scheme. It
is now generally accepted  \cite{debate} that  the PDS scheme is
equivalent to the cut-off scheme if the chiral counting is done
properly to the same order.

\begin{figure}[hbt]
%\centerline{\protect \epsfig{file=eft-fig3.eps}}
\centerline{\epsfig{file=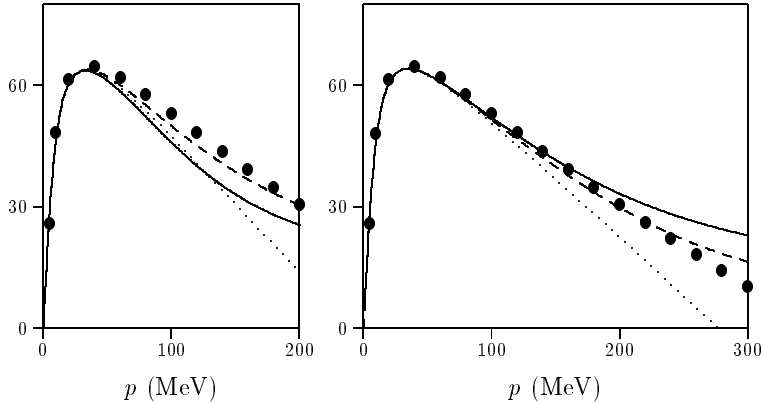, width=10cm}}
\caption[phase]{\protect \small
The $np$ $^1S_0$ phase shift (degrees) vs. the center-of-mass (CM)
momentum $p$. We show the predicted results both without (left) and
with (right) the pion field. For each case, our NLO results are
given for two extreme values of the cutoff: The solid curve
represents the lower limit, $\Lambda= 100$ MeV (without the pion)
and $200$ MeV (with the pion), and the dotted curve the upper
limit, $\Lambda= 300$ MeV (without the pion) and $500$ MeV (with
the pion). The ``experimental points" (obtained from the Nijmegen
multi-energy analysis) are given by the solid circles. Our theory
with $\Lambda=150$ MeV (without the pion) and $\Lambda= 250$ (with
the pion) are shown by the dashed line to show that the theory is
in almost perfect agreement with the data up to $p \lsim 200$ MeV.
}\label{fig2}
\end{figure}

The calculation with a cutoff to NLO in the irreducible vertex (or
potential) without any additional fiddling is fully consistent
\footnote{Furthermore it is so simple that it makes one wonder what
all this fuss about the PDS and other schemes is about.}. Figure
\ref{fig2} shows that the phase shifts come out very well, fairly
independently of the cutoff, up to $p\sim 100$ MeV if the pion is
present and to $p\sim 70$ MeV if the pion is absent. Again the
presence of the pion as an explicit degree of freedom markedly
improves the result.
\subsubsection{$\bullet$ The goodness of EFT: Insensitivity to $\Lambda$}
\indent\indent
Although there is little dependence on the value of $\Lambda$, the
results are not entirely independent of it. This is expected since
the series is truncated in the way the cutoff procedure goes. The
reason is simply that the (chiral) counting in the scheme with the
cutoff is not fully consistent, with the error residing in higher
orders. That is, a genuine $\Lambda$ independence requires
including certain terms of the next order and we are not including
them here, so they represent the error committed. That the results
are quite insensitive to the cutoff value indicates however that
the error committed is small enough.
\begin{figure}[hbtp]
%\centerline{\protect \epsfig{file=eft-fig2.eps}}
\centerline{\epsfig{file=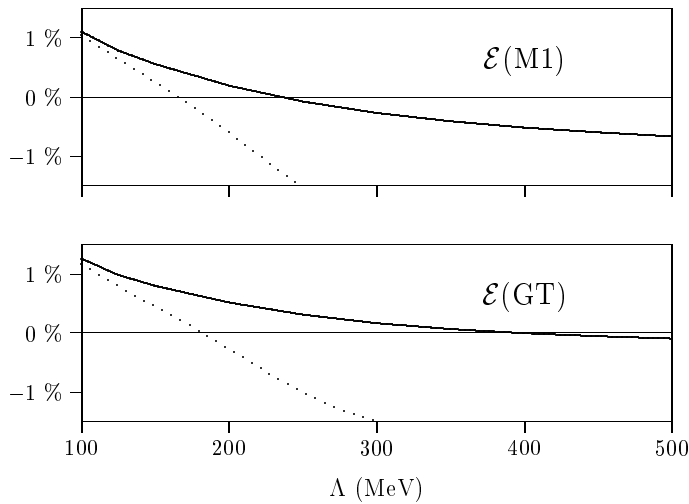, width=9cm}}
\caption{\protect \small
${\cal E}(\mbox{M1})$ (upper) and
${\cal E}(\mbox{GT})$ (lower) vs. the cutoff $\Lambda$.
The solid curves represents the NLO results with
pions and the dotted curves without pions.
\label{fig3}}
\end{figure}

Let us quantitatively examine whether and how our theory is a truly
{\it effective} effective theory. We do this using the electroweak
processes. This test has not yet been made by other workers in the
field. For this we should demand not only that the theory agree
with experiments but also that the result not depend sensitively on
the cutoff. We test the latter by looking at the following
quantities:
\bea
{\cal E}(\mbox{M1}) \equiv \frac{\calM_{\rm M1}^{\rm th}
 - \calM_{\rm M1}^{v18}}{\calM_{\rm M1}^{v18}},
\ \ \
{\cal E}(\mbox{GT}) \equiv \frac{\calM_{\rm GT}^{\rm th}
 - \calM_{\rm GT}^{v18}}{\calM_{\rm GT}^{v18}},
\eea
where $\calM_{\rm M1}^{\rm th}$ and $\calM_{\rm M1}^{v18}$ denote,
respectively, the M1 transition matrix element of our NLO
calculation and that of the Argonne $v_{18}$ potential~\cite{v18}, and
similarly for ${\cal E}(\mbox{GT})$. Here we are taking the Argonne
potential as ``experiment" since it is fitted to experiments. Since
these quantities are not the entire story that can be directly
compared with experiments (due to the exchange currents), this is
the best we can do to NLO. We see in Fig.\ref{fig3} that our
criterion is very well satisfied. This is the more so when the pion
is incorporated. {\it We see in a quantitative way how accounting
for ``relevant degrees of freedom" in effective theories improves
the result as well as the consistency of the theory.}
\subsubsection{$\bullet$ The chiral filter}
\indent\indent
I mentioned above that the advantage over other
consistent schemes (such as PDS) of the Weinberg counting
combined with the cut-off regularization we adopted is that the chiral filter
mechanism allows one to make certain {\it predictions,
not post-dictions} that are not
easily accessible to other schemes. The soft pion plays a
particularly prominent role in nuclear processes and this
predominance is visible in the scheme. Thus the importance of the
soft-pion exchange in nuclear isovector M1 transitions and in
nuclear axial-charge transitions was {\it predicted} before
experiments came to confirm it. A further forecast we can make is
that when the pion-exchange is suppressed for whatever reason,
chiral perturbation expansion will lose much, if not all, of
its predictive power and
one would be forced to adopt
a different strategy than the usual one. Indeed this
is exemplified by the inability of low-order chiral perturbation
calculations to explain the experimental cross section for the
$p+p\rightarrow p+p+\pi^0$ process. In other schemes, the
particular role of pions could be hidden, though not absent, and
one would have difficulty in ``seeing" it and hence in making
predictions that hinge on it.

As a specific example, let me take the radiative $np$ capture. The
precise agreement between the chiral perturbative calculation and
the experiment for the capture cross section obtained in \cite{pmr}
is the bench-mark in nuclear physics and the result was understood
due to the clear identification of the important role of the pion
exchange. In the recent calculation of the same process in the PDS
scheme \cite{SSW}, the authors emphasize the fact that the
pion-exchange graph in their scheme is {\it only one of several NLO
terms in the expansion, is scale- and scheme-dependent
 and hence has no particular significance.} Since
the experimental cross section is used to fix one unknown NLO
dimension-6 counter term, this calculation is not a prediction.
Indeed the counter term so extracted turns out to be as large as the
terms explicitly calculated, and there is nothing which says that
higher order (nonanalytic) terms and counter terms will be
negligible in the scheme.

I should hasten to say that there is
nothing basically
wrong in this PDS calculation. It is just that it does not seem
as predictive. My personal opinion on this matter is this. The PDS
scheme is perhaps more consistent in the spirit of EFT than the
scheme of ref.\cite{pmr}. However by strict adherence to the
consistency of the counting, it misses an important piece of physics,
namely, the
short-range correlation that is present in nuclear wave functions.

So what is this short-range correlation (SRC)? As far as I can see,
this is not understood within the framework of
EFT. It is an effect of short-distance physics that is ``integrated
out" from the EFT, so in principle if the expansion of the EFT
makes sense, the SRC effect presumably is lodged in the
parameters of the EFT. In this scheme, then, the question is: How
far does one have to go in the expansion to capture the essence of
SRC physics? In the nuclear physics community, the SRC appears in the wave
function as a sort of ``black box" determined by looking at
experiments. In \cite{pmr}, the SRC so invoked is found to suppress {\it all}
contact terms {\it in the current}~\footnote{Note that contact terms in
the irreducible diagrams contributing to the potential are not suppressed by the
same mechanism as explained in \cite{MR91}. This is the modern understanding
of the chiral filter conjecture.}. The result shows that this is
consistent with the chiral filter,  the precise agreement providing
a support for this reasoning: The error incurred is found to be
negligible. It seems possible to justify this SRC mechanism by
resorting, along the line developed by Lepage \cite{lepage}, to the
OPE in the wave function to factorize the short-distance
(asymptotically free) part of the QCD interaction inaccessible to
chiral perturbation theory. By delegating the short-distance
physics to the wave function as we did, we ``enhance" the terms
which are allowed by chiral filter and ``suppress" those which are
not. This is why there is no predictive power for our ChPT when
soft-pion terms are suppressed. On the other hand, the PDS scheme
may allow a ``systematic calculation," but there is no reason to
expect that it would work well.
\section{Effective Field Theories For Dense
Matter}
\indent\indent
The strategy found to be so successful at low energy for
two-nucleon systems is being extended to three-body systems
\cite{3body} but it would require a quantum jump to get to nuclear
matter and denser matter. This means that there cannot be a
``straightforward" application of chiral perturbation theory in the
immediate future. All we can do for the moment is to resort to some
educated guesses and check simplifying assumptions we make against
nature to see whether our guesses pass the test. We eliminate wrong
guesses and adopt the right ones by trial and error.\footnote{My point is
that if we sit and worry about the enormous complications brought
in by many-body dynamics under extreme conditions, we will get
nowhere. The analogy would be a centipede who sits and worries how
it is going to march forward with so many legs potentially in the
way.} I will describe in the rest of this lecture how we move
forward in this direction and  meet with some success. 
This part is based on a series of
papers written recently \cite{FR96,SBMR,SMR98,FRS98}.

The basic idea is to combine two superbly effective field theories
available in the market via  a simple scaling assumption which is
known in the literature as ``BR scaling."  The two EFTs that I
will invoke are chiral Lagrangian field theory we now understand
well and Landau Fermi liquid theory for dense matter with a Fermi
sea which can be formulated as an effective field theory~\cite{pol}.
We know how the first works in dilute systems and how
the latter works in strongly correlated fermionic systems. Our aim
is to bridge the two in a way that can be applied to dense hadronic
matter.
\subsection{Nuclear Matter as a Chiral Fermi Liquid}
\indent\indent
The first question we have to address is: What is nuclear matter in
chiral Lagrangian field theory? One might address this question in
terms of a skyrmion with a large winding number $A$ (see \cite{CND}
for references) but so far very little has been learned from this.
It seems more profitable to start with a chiral Lagrangian with
nucleons and mesons incorporated, calculate the effective action in
a suitable chiral expansion and look for a (nontopological) soliton
 with a given baryon number $A$. Such a solitonic solution
would then describe the ground state of a system with baryon number
$A$ which could be closely resembling  a heavy nucleus in its ground state.
Lynn \cite{lynn} has argued, using a somewhat oversimplified model,
that such a system is a chiral liquid that gives a realistic
description of a droplet of nuclear matter. Although there is no
convincing derivation of the chiral liquid solution, the reasoning
is very plausible and I am going to assume that this conjecture is
correct. The next crucial step is to identify this chiral liquid
structure with the mean field solution of a phenomenologically successful
 effective chiral
action such as that written by Furnstahl et al \cite{furnstahl} in terms of an
expansion based on ``naturalness condition." The effective chiral
Lagrangian theory of Furnstahl et al for nuclear matter is a
generalization of Walecka model~\cite{walecka}, so we arrive at the first
conclusion that the chiral liquid structure can be identified with
the mean field solution of a Walecka-type model.
\subsection{Effective Chiral Lagrangian with BR Scaling}
\indent\indent
I will next argue that when BR scaling is implemented into an
effective chiral Lagrangian, the mean-field theory of this
Lagrangian corresponds to Walecka mean field theory {\it extended}
a la Furnstahl et al.
\subsubsection{$\bullet$ BR scaling}
\indent\indent
It is highly likely that when baryon matter is present, the vacuum
itself on which excitations take place is modified. This is so
since the quantity that characterizes the vacuum, the quark
condensate $\la \bar{q}q\rangle$, must change in the presence of
matter as indicated by such models as Nambu-Jona-Lasinio model (see
\cite{CND}). We may consider writing an effective Lagrangian in a
space that contains a matter of density $\rho$ directly in terms of
the parameters defined in that space. The parameters then must
change as the density changes. One can think of the parameters
depending on some field variables that change according to the
change of the condensate and hence indirectly according to density.
We will then assume that the effective Lagrangian defined in that
space characterized by a particular field value (or density) be
dictated by the symmetries of QCD. In \cite{BR}, assuming the
relevant symmetries to be the chiral symmetry and the scale
symmetry of QCD phrased in the form of large $N_c$ theory, i.e.,
skyrmion, the mass parameters and the (pion decay) coupling
constant of the effective theory are found to scale as
\bea
\frac{M^\star_N}{m_N}\approx\frac{m_\omega^\star}{m_\omega}
\approx\frac{m_\rho^\star}{m_\rho}
\approx \frac{m_\phi^\star}{m_\phi}\approx \frac{f_\pi^\star}{f_\pi}
\equiv \Phi.\label{BRscaling}
\eea
Here the $\star$ represents in-medium quantity, $M^\star_N$ a
scaling nucleon mass unaffected by the pion cloud (the ``Landau
effective mass'' of the nucleon $m_N^\star$ will be defined later),
$\omega$ and
$\rho$ are respectively the isoscalar vector meson and the
isovector vector meson degrees of freedom and $\phi$ an isoscalar
scalar meson field which is a chiral singlet effective in
nuclear matter at a mass $\sim 500$ MeV.

\subsubsection{$\bullet$ Chiral symmetry and  Walecka model}
\indent\indent
It has often been asked why Walecka model~\cite{walecka}
works. In fact, Cohen et al~\cite{cohenQCDS} gave a QCD sum rule justification
of the large vector and scalar mean fields found in the Walecka-type models.
Contrary to what some people have claimed, the success of the model is
{\it not}
due to any of the following: (1) renormalizability, (2) relativity, (3)
accuracy of the mean field at large density.

The theory is an effective theory and {\it not} a fundamental theory,
so neither does renormalizability have any
raison d'\^etre nor should it be valid at asymptotic density at which the model
-- unless fundamentally changed -- should be irrelevant.
Since heavy-baryon (nonrelativistic) chiral
perturbation treatment (as given below) successfully reproduces the result of
Walecka-type models, relativity is clearly not an essential ingredient
for the success either.

So why does Walecka model work?

I suggest the reason is that it is consistent with chiral symmetry, as
explained in \cite{BR96,ritzi}. To see this, suppose we write an effective
Lagrangian in heavy-fermion formalism with {\it all} except nucleon fields
integrated out. The Lagrangian will consist of the usual nucleon bilinear
term and multi-Fermion interactions, $(\bar{N}\cdots N)^n$ with $n\geq 2$
with various covariants and derivatives appearing so that chiral symmetry
is preserved. In particular the four-Fermi interactions of the form
\bea
\bar{N} \Gamma^\alpha N \bar{N} \Lambda_\alpha N
\eea
where $\Gamma^\alpha=\Lambda^\alpha=1$ and $\Gamma^\alpha=\Lambda^\alpha
=\gamma^\alpha$ have respectively scalar meson (``$\sigma$'') and isoscalar
vector meson (``$\omega$'') quantum numbers which would, when suitably
treated in mean field, give rise to the same effects as the $\sigma$ and
$\omega$ fields in a Walecka-type theory. It is clear from this argument that
the scalar is not the sigma of the linear sigma model. It is a chiral
singlet and hence safe from the disaster in nuclear matter that the $\sigma$ of
the linear sigma model produces.

The most reasonable way of interpreting the scalar is that it is a
``quarkonium'' component of the dilaton that figures in the trace
anomaly~\cite{beane}. This is therefore a chiral singlet.
In free space, it must be massive $\sim 700$ MeV but go down to $\sim
500$ MeV at nuclear matter density.
This scalar must ultimately
join the pions at the chiral phase transition to make up the
quartet of the $O(4)$ chiral symmetry. This could occur via
Weinberg's mended symmetry~\cite{weinbergmend,beane}. Thus as density increases
toward the chiral symmetry restoration point, the mass of the scalar must drop.
BR scaling (\ref{BRscaling}) prescribes how the mass could drop.

The vector meson $\omega$ is also a chiral singlet. According to BR
scaling, its mass must drop as a function of density. How does this happen?
The way to see that the mass dropping is ``universal'' as in (\ref{BRscaling})
is that it is essentially a scalar tadpole acting on all hadrons in the way
that the Nambu-Jona-Lasinio model would describe, namely through the
dynamically
generated mass of a constituent quark making up the hadrons. It is the
higher dimension multi-Fermi operators present in the effective Lagrangian
that would generate the tadpole mechanism.
\subsubsection{$\bullet$ The chiral liquid as Landau Fermi liquid}
\indent\indent
If we assume that we have a chiral liquid defined with the Fermi momentum
$k_F$, then excitations around the liquid ground state can be treated again
as an effective field theory as described beautifully in \cite{pol}. This will
give essentially Landau Fermi liquid theory, with two fixed points, one
the effective mass of the nucleon $m_N^\star$ which we will identify as
Landau effective mass, and the other, Fermi liquid interactions
${\cal F}$ between quasiparticles.

The next important argument we need can be borrowed from Matsui~\cite{matsui}
who showed that Walecka theory for nuclear matter can be mapped to Landau
Fermi liquid theory with the vector-meson-mediated interaction related
to the Landau parameter $F_1$. Since this Landau parameter governs also the
Landau effective mass of the nucleon, the scaling of the nucleon mass must
be related {\it mainly}
to the vector degree of freedom~\footnote{Note that this differs from the
usual Walecka model where the nucleon mass shift comes from the
VEV of the $\phi$ field.} . With BR scaling this means that
the scaling of all hadrons must be governed by the Fermi-liquid fixed-point
parameter $F_1$.

Our conclusion then is: chiral Lagrangian with BR scaling, via its
identification with Walecka mean field theory, can be mapped to
Landau Fermi liquid fixed point theory. For details of the
reasoning, I refer to \cite{FR96,SBMR,SMR98,FRS98}.
\subsection{Predictions at Normal Nuclear Matter Density}
\indent\indent
Here we shall establish a crucial relation between
BR scaling and the Landau parameter $F_1$ and give some
predictions based on the relation.
\subsubsection{$\bullet$ Nuclear matter}
\begin{figure}[tbh]
\setlength{\epsfysize}{2.5in}
\centerline{\epsffile{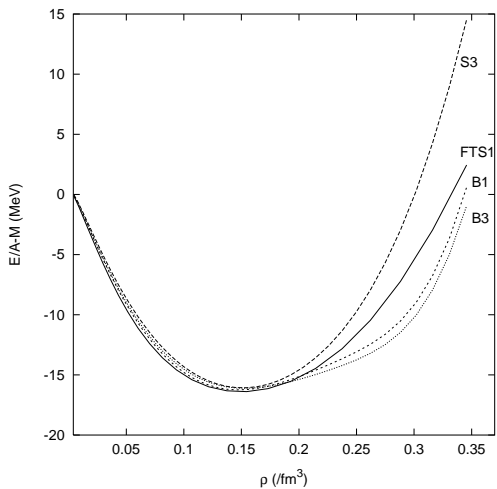}}
\caption[good]{$E/A-M$ vs. $\rho$
for this model ($S3$) compared with the elaborate model (labeled as
$FTS1$) of Furnstahl et al~\cite{furnstahl}. $B1$ and $B3$ include
higher polynomial terms in the Lagrangian (\ref{model}) that
represent fluctuations beyond the mean field. }
\label{sbmr}
\end{figure}

\indent\indent
We learned from the chain of arguments developed above that
in order to describe nuclear matter, all we need to do is to write the
simplest form of Walecka model with masses and coupling constants affixed with
a star and let the masses scale according to (\ref{BRscaling}). The Lagrangian
is
\bea
\L&=&\bar{N}[\gamma_\mu (i\del^\mu-g_v^\star (\rho )
\omega^\mu )-M^\star (\rho )
+h^\star \phi ] N\label{model}\\
& &+\frac12[(\del\phi )^2-m_s^{\star 2}(\rho )\phi^2]-\frac14 F_\omega^2
+\frac12 m_\omega^{\star 2}(\rho )\omega^2\nonumber
\eea
where $N$ is the nucleon field, $\omega_\mu$ the isoscalar
vector field and $\phi$ an isoscalar scalar field. Pseudoscalar fields can
be suitably incorporated as done later but they do not figure in the symmetric
nuclear matter that we are concerned with here. The
scaling behavior of the constants $g_v$ and $h$ is left arbitrary. It was shown
in \cite{SBMR} that in the mean field, this Lagrangian with the
BR scaling (\ref{BRscaling}), a suitable scaling of the constant $g_v$
and no scaling of $h$ gives a surprisingly good
description of the ground state with a compression modulus well
within the accepted value $200\sim 300$ MeV. The result is given in
Fig.\ref{sbmr}. The scaling $\Phi (\rho_0)=0.78$
needed here is forced on us, as described below, by QCD sum-rule calculations
and orbital gyromagnetic ratios in heavy nuclei.
Contrary to the worry expressed
by some people, when properly interpreted, the
density dependence of the parameters of the Lagrangian does not spoil any
thermodynamic consistency of the theory~\cite{SMR98}.
\subsubsection{$\bullet$ BR scaling $\Phi$ and Landau parameter $F_1$}
\indent\indent
From Galilean invariance (or Lorentz invariance) follows the Landau effective
mass of the nucleon
\bea
\frac{m_N^\star}{m_N}=1+F_1/3=\left(1-\tilde{F}_1/3\right)^{-1}\label{landau}
\eea
where $\tilde{F}=(m_N/m_N^\star) F$. On the other hand, the chiral Lagrangian
with BR scaling gives in tree order
\bea
\frac{m_N^\star}{m_N}=\Phi(1+F^\pi_1/3)=\left(\Phi^{-1}-
\tilde{F}^\pi_1/3\right)^{-1}\label{chiral}
\eea
where $F_1^\pi$ is the contribution from the pion to the Landau $F_1$
parameter. Now comparing (\ref{landau}) and (\ref{chiral}), we see
that
\bea
\tilde{F}_1-\tilde{F}_1^\pi=3(1-\Phi^{-1}).
\eea
It seems reasonable to assume  that the 
contribution to $F_1$ from the $\omega$ channel saturates the
rest. Then we have
\bea
\tilde{F}_1^\omega=3(1-\Phi^{-1}).
\eea
This is a crucial relation between BR scaling and the $\omega$ degree
of freedom. It
shows that the nucleon effective mass is primarily governed by the $\omega$
degree of freedom as mentioned above.

Calculation of the response of a nucleon sitting
on top of the Fermi sea to
electromagnetic current gives the gyromagnetic ratio that agrees
precisely with the Landau-Migdal formula~\cite{migdal}
\bea
g_l=\frac{1+\tau_3}{2}+\delta g_l
\eea
with
\bea
\delta g_l=\frac 16 (\tilde{F}_1^\prime-\tilde{F}_1)\tau_3
\eea
where $F_1^\prime$ is the isovector counterpart to $F_1$.
The chiral Lagrangian with BR scaling gives in tree order
\bea
\delta g_l=\frac 49 \left[\Phi^{-1}-1 -\tilde{F}_1^\pi\right]\tau_3.
\label{deltagl}
\eea
Since the $F_1^\pi$ is completely known by chiral symmetry,
given a $\delta g_l$ from experiment, one can determine $\Phi$
at nuclear matter density. We can turn the problem around and calculate
$\delta g_l$, given $\Phi (\rho_0)$. We shall do the latter by noting
that $\Phi (\rho_0)$ can be calculated from Gell-Mann-Oakes-Renner
formula for the pion in medium or from QCD sum-rule calculations of
the vector meson mass in medium and using BR scaling
(\ref{BRscaling}). The two ways give the same answer. Take the QCD
sum-rule value~\cite{QCDSR}
\bea
\Phi (\rho_0)=0.78\pm 0.08.
\eea
With this value and the known $F_1^\pi$ at $\rho=\rho_0$
(for which the Landau mass comes out to be $m_N^\star (\rho_0)/m_N
\approx 0.70$), we get
\bea
\delta g_l (\rho_0)=0.227\tau_3.
\eea
This agrees precisely with the experimental value obtained 
for the proton from
giant dipole resonances in heavy nuclei
\bea
\delta g_l^{(exp)}=0.23\pm 0.03.
\eea
\subsubsection{$\bullet$ Fluctuations around nuclear matter ground state}
\indent\indent
Pseudo-Goldstone octet bosons fields can be readily restored in the
Lagrangian (\ref{model}) in a way consistent with chiral symmetry.
Their fluctuations will describe their properties in the medium
{\it defined} by the scaling parameter $\Phi$. For example,  kaonic
fluctuations would have the form
\bea
{\cal L}^{eff}=\frac{-6i}{8{f_\pi^\star}^2}\overline{K}\del_t K
N^\dagger N+\frac{\Sigma_{KN}}{{f_\pi^\star}^2} \overline{K}K
\eea
where $\Sigma_{KN}$ is the $KN$ sigma term. In tree order (i.e.,
mean field), $\overline{N}N\approx N^\dagger N\approx \rho$.  The
potential felt by the kaon in the background of nuclear matter is
given by
\bea
V_{K^{\pm}} &=& \pm \frac{3}{8{f_\pi^\star}^2}\rho,\\
S_{K^\pm}&=&-\frac{\Sigma_{KN}}{{f_\pi^\star}^2}\rho_s
\eea
where $\rho=\la N^\dagger N\rangle\approx\rho_s=\la
\overline{N}N\rangle$. At nuclear matter density, we can identify
these results as one-third of the corresponding potentials for
nucleons, so we can write
\bea
V_{K^\pm}\approx \pm \frac 13 V_N
\eea
and
\bea
S_{K^\pm}\approx \frac 13 S_N.
\eea
Phenomenology in Walecka-type mean-field theory gives
$(S_N-V_N)\lsim -600\ {\mbox{MeV}}$ for $\rho=\rho_0$.
This leads to
the prediction that at nuclear matter density
\be
S_{K^-}+V_{K^-}\lsim -200\ {\mbox{MeV}}.
\ee
This is consistent with the result coming from the analysis in K-mesic atoms
$\left(S_{K^-}+V_{K^-}\right)^{K-atom}\sim -200 {\mbox{MeV}}$.
This order of the mass shift in the kaon mass is also seen in heavy-ion
experiments and found
to lead to an interesting consequence on the formation of neutron stars and
light-mass black holes. This issue is discussed by Li, Lee and
Brown~\cite{LLB}.
\subsubsection{$\bullet$ Strongly enhanced axial-charge transitions in 
heavy nuclei}
\indent\indent
As mentioned at the beginning, Warburton's experiments and analyses
on weak axial-charge transitions~\cite{warburton}
confirmed one of the chiral filter predictions. There was also a surprise
there which excited quite a few
theorists. Warburton found that not only was there
a strong mesonic contribution to the axial-charge matrix element but 
even more interestingly, the
enhancement over the single-particle matrix element in heavy nuclei (such
as lead) was 100 \%. It turns out 
that this strong enhancement in dense nuclei can be simply explained by
BR scaling as proposed by Kubodera and Rho~\cite{KR}.

Consider, following Warburton, the quantity $\epsilon_{MEC}$ which is the
ratio of the {\it measured} axial-charge matrix element over the 
{\it theoretical} single-particle
matrix element. Warburton's result for the lead region
was that
\bea
\epsilon_{MEC}^{exp}\approx 1.9\sim 2.0.
\eea
Now using our chiral Lagrangian with BR scaling in the tree order, 
we get the very simple formula~\cite{FRS98}
\bea
\epsilon_{MEC}^{th}\approx \Phi^{-1} (1+{\cal R})\label{epsilon}
\eea
where ${\cal R}$ is the ratio of the (soft-pion dominated)
meson-exchange current matrix 
element over the single-particle matrix element which for $\rho=\rho_0$
comes out to be $0.56\sim 0.61$ (this is a quite reliable range 
protected by the chiral filter mechanism~\cite{FRS98,pmrPR}). 
For the lead region in question,
we may take $\rho\approx \rho_0$. We find for the range involved for
${\cal R}$,
\bea
\epsilon_{MEC}^{th} (\rho_0)= 2.0\sim 2.1.
\eea
In my opinion, this is as strong a confirmation of the notion of BR scaling
as the orbital gyromagnetic ratio discussed above.
\subsection{Predictions For Higher Density $\rho >\rho_0$}
\indent\indent
We have so far seen that used at the mean field level, the chiral effective
Lagrangian with BR scaling can reasonably well describe nuclear properties at
nuclear matter density. The question is: Can one extend the treatment to higher
densities at which various phase changes such as kaon condensation, color
superconductivity, chiral restoration etc. could take place? In \cite{LLB},
it was argued that kaon condensation could take place 
at $\rho\lsim 3\ \rho_0$ within the framework. What about the others? For this,
one must go beyond the mean field with the effective Lagrangian but so far
there is no systematic formulation of going higher order with the Lagrangian
with BR scaling.

The mean field arguments are expected to work provided quasiparticle notions
apply to the hadrons relevant to the process. For instance, BR
scaling has been successfully applied to the CERES dilepton
processes~\cite{LKB} which
involve densities of $2\sim 3$ $\rho_0$ and temperatures of $\gsim 100$ MeV.
Here the principal mechanism that is invoked is the mass shift of the $\rho$
meson with the width playing a secondary role.
The reasoning here is that when one fluctuates around the background
at a given density with scaling masses and other parameters, most of the strong
correlations are included in those effective parameters and the residual
interactions on top of the given background are {\it weak}
so that a weakly-interacting local field
description is valid. Thus it may be that
viewed from BR scaling, widths do not destroy the
quasiparticle picture.

On the other hand, if one were to build correlations starting
from the matter-free
background, one is in a strong-coupling regime and
it seems inevitable that interactions broaden the width of
the hadrons involved, in particular, of the $\rho$ meson.
It is found that the medium-broadened
width of the vector meson can also explain reasonably well the CERES
data~\cite{wambach}.

Is there any connection between the mass-shifted
vector meson defined at a density $\rho\neq 0$ and the broadened-width
vector meson built on the $\rho=0$ background? My conjecture is that there is
a dual description of the same process in terms of different languages,
similar to the CCP discussed in the first lecture. Here the languages dual to
each other are the ``partonic'' picture of ref.~\cite{LKB} (with BR-scaling
hadrons) and the
``hadronic'' picture of ref.~\cite{wambach} (with many-body interactions).
This is somewhat like the
quark-hadron duality one sees in heavy-light meson decay
processes~\cite{quark-hadron}.
How this dual structure can work in the case of the CERES dilepton process
is explained in a recent paper co-authored by the two schools --
BR scaling and dynamical width~\cite{dual}.

\section*{Acknowledgments}
\indent\indent
This lecture is based on a series of work I have done over the
years with my collaborators, in particular, Gerry Brown, Marc
Chemtob, Jean Delorme, Bengt Friman, Kuniharu Kubodera, Kurt
Langfeld, Chang-Hwan Lee, Dong-Pil Min, Byung-Yoon Park, Tae-Sun
Park, Chaejun Song, Vicente Vento, Ismail Zahed. This lecture was
prepared at Korea Institute for Advanced Study (KIAS) and I am
grateful to Chung Wook Kim, Director of KIAS, for his support and
encouragement.

%\section*{Appendix}


\begin{thebibliography}{99}
\bibitem{gegelia2} J. Gegelia, ``Are pions perturbative in effective
field theory?", nucl-th/9806028.
\bibitem{ravndal} Most recent treatment of this matter in EFT is given
by X. Kong and F. Ravndal, ``Proton-proton scattering lengths from
effective field theories,'' nucl-th/9811076
\end{thebibliography}
\end{document}